\documentclass[%
 reprint,
 amsmath,amssymb,
 aps,
]{revtex4-2}

\usepackage{graphicx}
\usepackage{dcolumn}
\usepackage{bm}
\usepackage{color}

\begin{document}

\preprint{APS/123-QED}

\title{Protocol to evaluate the viscoelastic response of a polymer suspension to an active agent via oscillatory shear rheometry}

\author{Kai Qi$^{1,2}$}
\email{Contact author: kqi@mail.sim.ac.cn}

\author{Qingzhi Zou$^{1,2}$}

\author{Ignacio Pagonabarraga$^{3,4}$}
\email{Contact author: ipagonabarraga@ub.edu}

\affiliation{
$^{1}$2020 X-Lab, Shanghai Institute of Microsystem and Information Technology, Chinese Academy of Sciences, Shanghai 200050, China \\
$^{2}$Center of Materials Science and Optoelectronics Engineering, University of Chinese Academy of Sciences, Beijing 100049, China \\
$^{3}$Departament de F\'isica de la Mat\`eria Condensada, Universitat de Barcelona, C. Mart\'i Franqu\`es 1, 08028 Barcelona, Spain\\
$^{4}$Universitat de Barcelona Institute of Complex Systems (UBICS), Universitat de Barcelona, 08028 Barcelona, Spain}

\begin{abstract}
Microorganisms inhabit viscoelastic environments, where their locomotion can deform polymers and trigger local complex viscoelastic responses. However, a systematic approach to quantify such responses remains lacking. Here, we propose a protocol that maps the shear effect induced by an active agent to oscillatory shear rheometry. The central idea is to establish a correspondence between the mean shear rate generated by swimming and that produced by an oscillating plate. In this mapping, the swimming velocity and active stress are translated into an effective oscillation frequency and strain amplitude. The resulting viscoelastic response can then be evaluated by standard oscillatory rheometry. The protocol is validated using lattice Boltzmann simulations of a squirmer embedded in polymer solutions. Our framework is generic and can be naturally extended to active microrheology, providing a pathway to quantify swimmer-induced viscoelasticity.
\end{abstract}

\maketitle


Active agents moving through polymeric media generate nonequilibrium conditions where swimmer‐driven flows, polymer deformation, and viscoelastic stresses are tightly coupled \cite{li2021microswimming}. Swimmers both probe and reshape the surrounding rheology: their self-propulsion induces local shear, stretches nearby polymers, and elicits nonlinear viscoelastic responses that feed back on their motion. These reciprocal interactions significantly alter swimming kinematics, motility, and transport, producing phenomena governed jointly by swimmer type and fluid rheology \cite{patteson2015running,Gomez2016JanusParticle,Narinder2018Achiral,liu2021viscoelastic,qi2020rheotaxis,qi2025unravel,qi2020enhanced,Theers2018Clustering,Qi2022ActuveTurbulence,Qin2015Flagellar,Zoettl2019Enhanced,mo2023hydrodynamic,zhang2023locomotion,Tung2017sperm,Corato2021Spontaneous,Corato2025Janusdisk,Bozorgi2014Nonlinear,Hemingway2015Active,Spagnolie2013Locomotion,Tung2017sperm,gonzalez2025morphogenesis,liao2023viscoelasticity,Li2016Collective}. Examples include enhanced bacterial speed and trajectory persistence due to polymer stretching \cite{patteson2015running}, increased rotational diffusion and persistent circular motion of active colloids in viscoelastic media \cite{Gomez2016JanusParticle,Narinder2018Achiral}, and the emergence of large-scale vortices in confined DNA solutions \cite{liu2021viscoelastic}. Despite substantial progress in understanding how complex fluids modify swimmer behavior, the inverse problem—quantifying how a swimmer’s self-generated flow perturbs and reorganizes the surrounding polymeric medium—remains largely unsolved. Existing rheological techniques assessing fluid viscoelasticity fall into two categories. Macroscopic oscillatory and interfacial rheometry apply externally controlled stresses to extract linear or nonlinear viscoelastic moduli~\cite{charlton2019regulating}. At the microscale, passive microrheology infers linear viscoelastic moduli from the thermal fluctuations of embedded colloidal tracers \cite{mason1995optical,mukhopadhyay2001micro}, while active microrheology drives micrometer-sized probes through the medium using external forces~\cite{pecora2013dynamic,stewart2015artificial,ganesan2016associative,sretenovic2017early,galy2012mapping,zrelli2013bacterial,du2001study}. In particular, Khair \textit{et al.} established the theoretical framework for oscillatory active microrheology by relating frequency-dependent probe-induced deformations to the complex viscosity and linear viscoelastic moduli \cite{khair2005microviscoelasticity}. The rheological approaches have unraveled rich dynamical behavior in active systems \cite{mizuno2007nonequilibrium,gagnon2020shear,gupta2021rheology,knevzevic2021oscillatory,szakasits2019rheological,weber2022power}. For instance, active microrheology has revealed stiffening of cytoskeletal networks due to motor-generated filament tension \cite{mizuno2007nonequilibrium}, while oscillatory rheology has uncovered the nonmonotonic viscosity of microtubule-based active gels as they transition from fluid-like to solid-like states through the interplay of motor activity and external shear \cite{gagnon2020shear}. However, these methods rely on externally imposed stresses and thus cannot directly quantify the local viscoelastic response elicited by swimming, limiting insight into the mechanisms governing microswimmer behavior in viscoelastic media.

Here we introduce a mapping protocol that recasts the shear induced by a microswimmer into the language of classical oscillatory shear rheometry \cite{tseng2010linear,cifre2004linear}. By identifying a characteristic lateral distance at which the swimmer’s flow vanishes, we define an equivalent rheometric geometry in which the swimming speed and active stress yield an effective oscillation frequency and strain amplitude. Using lattice Boltzmann simulations \cite{lbm2001,Nash2008lbm,dunweg2009lattice,stratford2008parallel,pozrikidis1992boundary} of polymer suspensions under oscillatory shear and squirmer-generated flows, we demonstrate that this mapping enables direct evaluation of the dynamic moduli associated with swimmer activity. This framework provides a systematic route to quantify local swimmer-induced viscoelastic response and establishes a bridge between active-matter dynamics and rheological characterization.

\begin{figure}
  \centering
  \includegraphics[width=0.8\columnwidth]{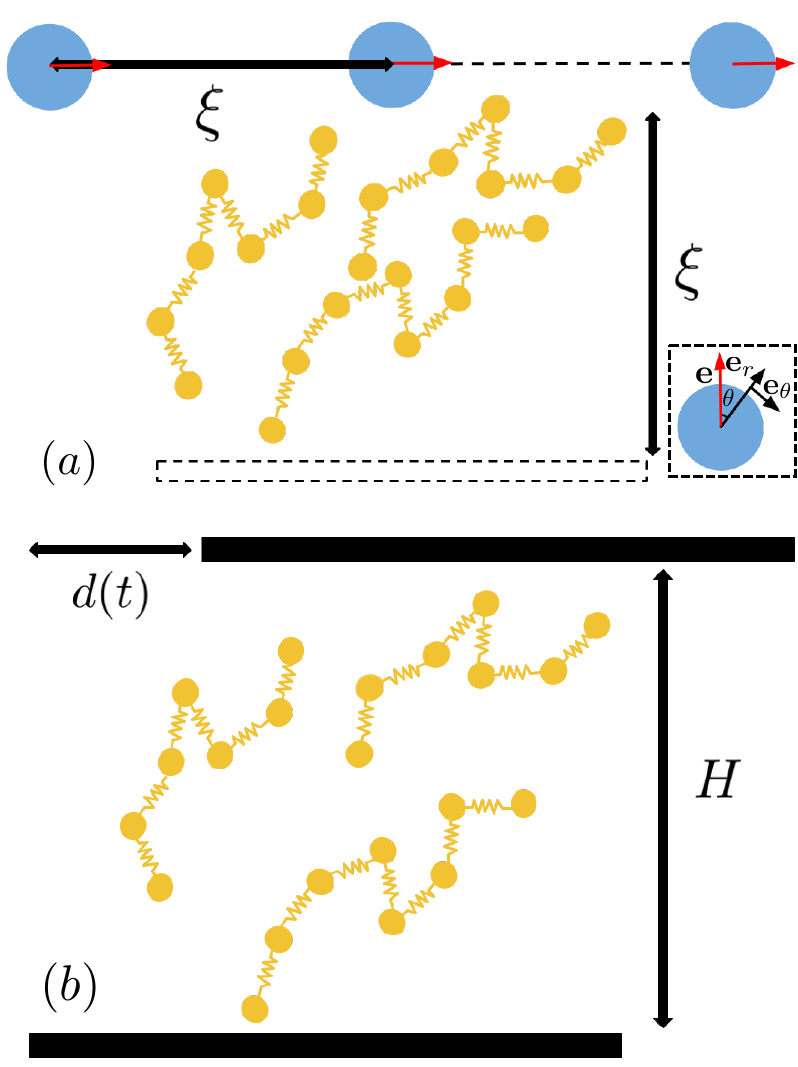}
    \caption{Schematic of polymers subjected to the shear flow generated by a microswimmer. The swimmer approaches and departs from the central polymers at a characteristic length scale $\xi$. At this distance, the perpendicular flow field has effectively vanished, introducing the equivalent of a static, imaginary wall at that location. The inset demonstrates the squirmer model. (b) Schematic of polymers embedded in an oscillatory shear flow. Two parallel walls are separated by a distance $H$. The bottom wall remains fixed, while the top wall oscillates with frequency $\omega$ and displacement $d(t)$.}
  \label{oscil}
\end{figure}

As illustrated in Fig.~\ref{oscil}(a), a swimmer initially positioned at a large lateral distance $\xi$ from the observation point moves horizontally from left to right. At this initial separation, the central polymers experience negligible flow. As the swimmer approaches, the local flow induces progressive polymer deformation, reaching a maximum when it passes directly through the center, and subsequently relaxes as the swimmer departs. This sequence is analogous to the oscillatory shear experiment (Fig.~\ref{oscil}(b)), where the wall executes a half-period oscillation, imparting a comparable shear deformation to the polymers. For the mapping to be valid, two conditions are required. (1) A characteristic length $\xi$ at which the swimmer’s flow effectively vanishes, serving as the position of an imaginary, static wall perpendicular to the swimmer’s trajectory. Laterally, $\xi$ plays the role of a quarter-wavelength of the oscillation, allowing the effective frequency $\omega$ to be estimated with the swimming speed $U_0$ as $\omega=\pi U_0/2\xi$. (2) The central idea is to establish a correspondence between the mean shear rate induced by swimming and that generated by the oscillating plate. With this mapping, standard oscillatory shear rheometry can be used to quantify the local viscoelastic response to swimmer activity by translating the swimming velocity and active stress into an effective oscillation frequency and strain amplitude. We verify the protocol by directly comparing computational oscillatory shear experiments with squirmer-induced flows via lattice Boltzmann simulations.

As shown in Fig.~\ref{oscil}(b), the oscillating top wall generates a time-dependent shear strain in the fluid 
\begin{equation}
    \gamma_w(t)\equiv \frac{d(t)}{H} =\gamma_0 {\rm sin}(\omega t), \label{shear_strain}
\end{equation}
where $t$ denotes time, $d(t)$ is the lateral displacement of the top wall, $H$ is the system height, $\gamma_{0}$ is the strain amplitude, and $\omega$ is the oscillation frequency. The corresponding shear rate is
\begin{equation}
    \dot{\gamma}_w(t)=\gamma_0 \omega {\rm cos}(\omega t). \label{shear_rate}
\end{equation}
In the fluid, the $xz$-component of the extra stress due to the presence of flexible polymers \cite{doi1988theory}  is
\begin{equation}
    \sigma^p_{xz}=-\frac{1}{V} \sum^N_{i=1}\sum^N_{j>i} r^m_{ij,x} f^{m}_{ij,z}, \label{stress_xz}
\end{equation}
where the sum runs over all monomers $N$, $V$ is the system volume, and $r^{m}_{ij}$ and $f^{m}_{ij}$ denote the relative distance and interaction force between monomers $i$ and $j$, respectively. The force $f^{m}_{ij}$ includes contributions from the FENE bond \cite{Kremer1990Entangled} and the short-range soft-sphere repulsion (see Supplemental Material, SM). Since each monomer is modeled as an inertialess point particle (see SM), the contribution from monomer kinetic energy to the stress tensor is neglected. Under oscillatory shear, a phase shift emerges between the polymeric stress component  
\begin{equation}
    \sigma^p_{xz}=\sigma_0{\rm sin}(\omega t+\delta), \label{stress_xz_oscil}
\end{equation}
and the applied shear strain $\gamma$, with stress amplitude $\sigma_{0}$ and phase angle $\delta$. The viscoelastic response of the fluid is characterized by two dynamic moduli: (i) the storage modulus $G'$, quantifying elastic energy stored by polymers and reflecting solid-like behavior, and (ii) the loss modulus $G''$, measuring viscous energy dissipation and reflecting liquid-like behavior. The moduli are defined as
\begin{equation}
    \frac{\sigma^p_{xz}}{\gamma_0} = \frac{\sigma_0}{\gamma_0}{\rm sin}(\omega t+\delta)
                                   = G'{\rm sin}\omega t + G''{\rm cos}\omega t, \label{moduli_def} 
\end{equation}
i.e., $G'\equiv \sigma_0/\gamma_0{\rm cos}\delta$ and $G''\equiv \sigma_0/\gamma_0{\rm sin}\delta$.
The corresponding loss tangent is 
\begin{equation}
    {\rm tan} \delta =G''/G'. \label{loss_tan}
\end{equation}
The rheological state of the fluid is determined by the phase angle $\delta$. For a Hookean solid, $\delta = 0$, while for a Newtonian fluid, $\delta = \pi/2$. Viscoelastic behavior corresponds to $0 < \delta < \pi/2$. In particular, when $G' > G''$, the fluid exhibits predominantly elastic, solid-like behavior, whereas $G' < G''$ indicates predominantly viscous, liquid-like behavior \cite{dealy2012melt}.

\begin{figure}
  \centering
  \includegraphics[width=\columnwidth]{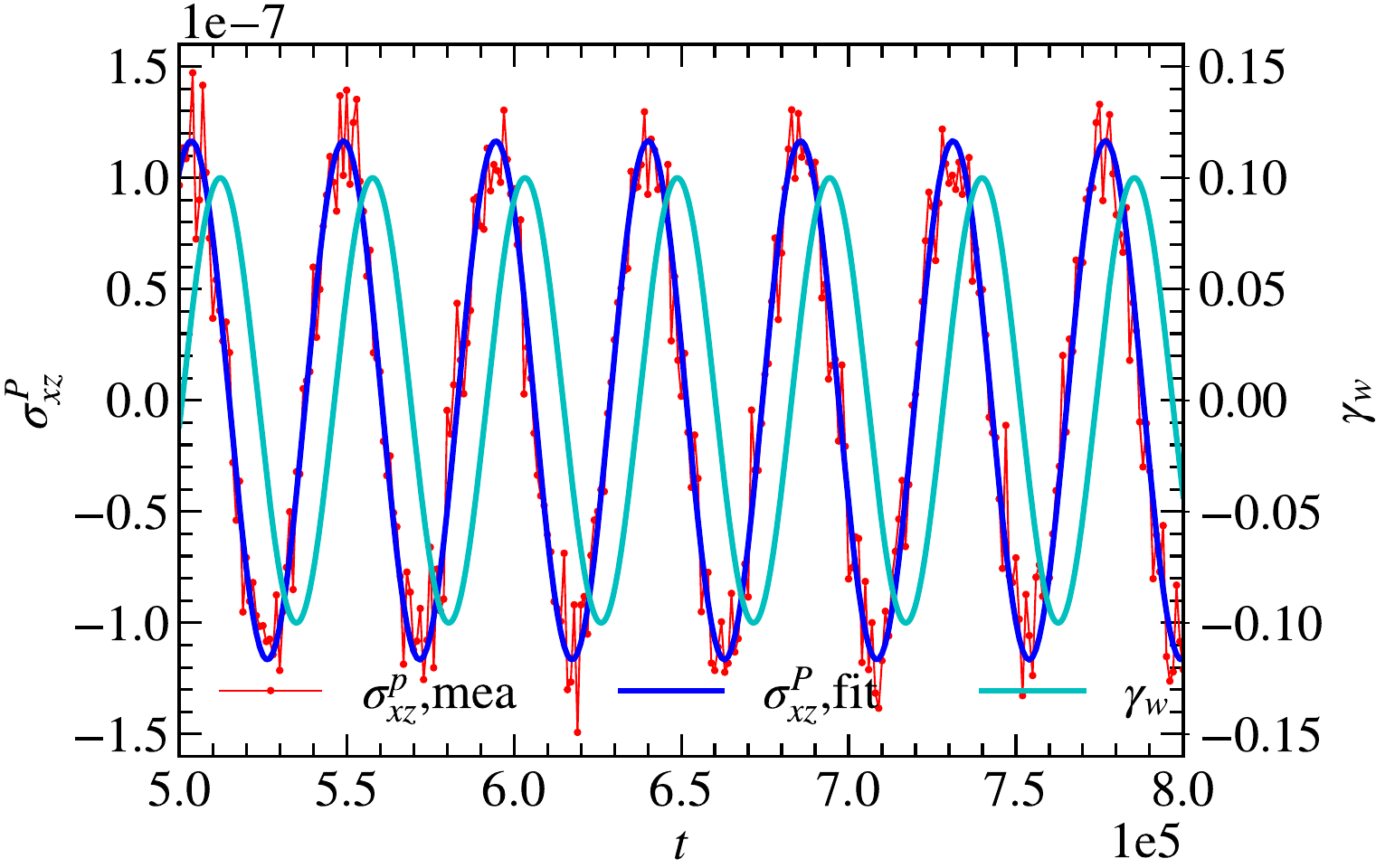}
    \caption{
        Time series of the imposed oscillatory shear strain $\gamma_w$ and the corresponding measured (red line) and fitted (blue line) polymeric stress components $\sigma^{p}_{xz}$, respectively.}
  \label{stress_tseries}
\end{figure}

To estimate the dynamic moduli, the stress amplitude $\sigma_{0}$ and phase angle $\delta$ are first extracted from the oscillatory shear stress response. The moduli are then directly obtained from their definitions. As an illustrative example, we consider a polymeric system confined between two parallel walls with box height $H = 38a$, containing $N_{p} = 50$ flexible polymers of length $N_{m} = 240$. The system is subjected to an oscillatory shear flow with frequency $\omega = 1.38 \times 10^{-4}\,\tau^{-1}$ and strain amplitude $\gamma_{0} = 0.1$ (Fig.~\ref{oscil}(b)). All quantities are expressed in lattice units with characteristic distance $a$ and time $\tau$ set to unity. Fig.~\ref{stress_tseries} shows the polymeric stress $\sigma^{p}_{xz}$, which oscillates with the imposed strain but with a clear phase lag. A least-squares fit~\cite{press1988numerical} yields $\delta = 1.2$ and $\sigma_0 = 1.17\times10^{-7}\,m/(a\tau^2)$. The corresponding moduli are $G' = 4.3\times10^{-7}\,m/(a\tau^2)$ and $G'' = 1.1\times10^{-6}\,m/(a\tau^2)$, giving a loss tangent $G''/G' = 2.55$. The intermediate phase angle and dominant viscous contribution indicate a viscoelastic yet predominantly fluid-like response.

To investigate the influence of microswimmer activity on the viscoelastic response of the fluid, we employ the squirmer model. As shown in the inset of Fig.~\ref{oscil}(a), the prescribed tangential slip velocity \cite{lighthill1952squirming,blake1971spherical,datt2017active,theers2016modeling,Theers2018Clustering,qi2020enhanced} on the spherical squirmer surface is
\begin{equation}
    {\bf u}_s=B_1 \sin(\theta)[1+\beta \cos(\theta)] \mathbf{e}_{\theta}, \label{u_sq}
\end{equation}
where $\theta = \arccos(\mathbf{e}\cdot\mathbf{e}_{r})$ is the polar angle between the radial unit vector $\mathbf{e}_{r}$ and the swimming direction $\mathbf{e}$, and $\mathbf{e}_{\theta}$ is the local tangent vector. The swimming speed $U_{0} = 2B_{1}/3$ is determined by the first squirming mode $B_{1}$ of the Legendre polynomial expansion. The dimensionless parameter $\beta$ characterizes the active stress: $\beta < 0$ for pushers, $\beta = 0$ for neutrals, and $\beta > 0$ for pullers. The squirmer swims horizontally from a large separation $\xi$, passes through the central polymer region, and departs, generating substantial polymer deformation—an effect analogous to the shear imparted during a half-period oscillation in oscillatory rheometry. The resulting flow field \cite{theers2016modeling} is
\begin{multline}
    {\bf v}(r,\theta)=\frac{B_1}{3} \frac{R^3}{r^3} (2{\rm cos}\theta {\bf e}_r + {\rm sin}\theta 
                      {\bf e}_{\theta})\\
                      - \frac{B_2}{2}\frac{R^2}{r^2}(3{\rm cos}^2\theta -1){\bf e}_r, 
                      \label{flow_field_squ_full}
\end{multline}
where $r$ is the distance to the center of the squirmer, $R = 3\,a$ denotes the squirmer radius, $B_2=B_1\beta$ is the second squirming mode, and the azimuthal symmetry is employed. As shown in the End Matter, the characteristic length scale for a squirmer is chosen as $\xi = 38\,a$, corresponding to a two-order-of-magnitude decay of the flow. The effective period can be thus estimated as 
\begin{equation}
    T=\frac{4\xi}{U_0} \label{period}
\end{equation}
and the corresponding frequency is 
\begin{equation}
    \omega=\frac{2\pi}{T}=\frac{\pi U_0}{2\xi}. \label{frequency}
\end{equation}

The squirmer-induced shear flow exhibits nonlinear variation near the central polymers. The mean shear rate generated by swimming is estimated by averaging the instantaneous shear rate over the spherical shell between $R$ and $\xi$, as given by 
\begin{equation}
    \langle \dot{\gamma}^2_s \rangle =\langle 2 \mathbf{D}:\mathbf{D} \rangle=\frac{9U^2_0 R}{4\xi^3} \left(4+\frac{18}{5}\beta^2\right), \label{mean_shear_rate_s}
\end{equation}
where $\mathbf{D}$ denotes the strain-rate tensor with components $D_{ij} = \frac{1}{2}\left( (\nabla\mathbf{v})_{ij} + (\nabla\mathbf{v})_{ji} \right)$ \cite{ladisa2005circumferential} (see End Matter). The corresponding mean shear rate due to wall oscillation at the same frequency is obtained as 
\begin{equation}
    \langle \dot{\gamma}^2_w \rangle =\frac{1}{T}\int^{T}_{0} \gamma^2_0 \omega^2 {\rm cos}^2(\omega t) dt=\frac{\pi^2 \gamma^2_0 U^2_0}{8\xi^2}. \label{mean_shear_rate_w}
\end{equation}
Equating the two shear rates yields the effective strain amplitude
\begin{equation}
    \gamma_0=\frac{3\sqrt{2}}{\pi} \left(\frac{R}{\xi}\right)^{\frac{1}{2}} \left(4+\frac{18}{5}\beta^2 \right)^{\frac{1}{2}}. \label{strain_amp_s}
\end{equation}
For $\xi = 38\, a$ and $R = 3\, a$, the effective strain amplitude increases monotonically with active stress, as shown in Fig.~S1. Thus, the influence of squirmer activity on the local viscoelastic response can be quantified by mapping the swimming speed and active stress onto the equivalent oscillation frequency and strain amplitude using Eqs.~\ref{frequency} and \ref{strain_amp_s}, followed by standard oscillatory shear rheometry.

\begin{figure}
  \includegraphics[width=\columnwidth]{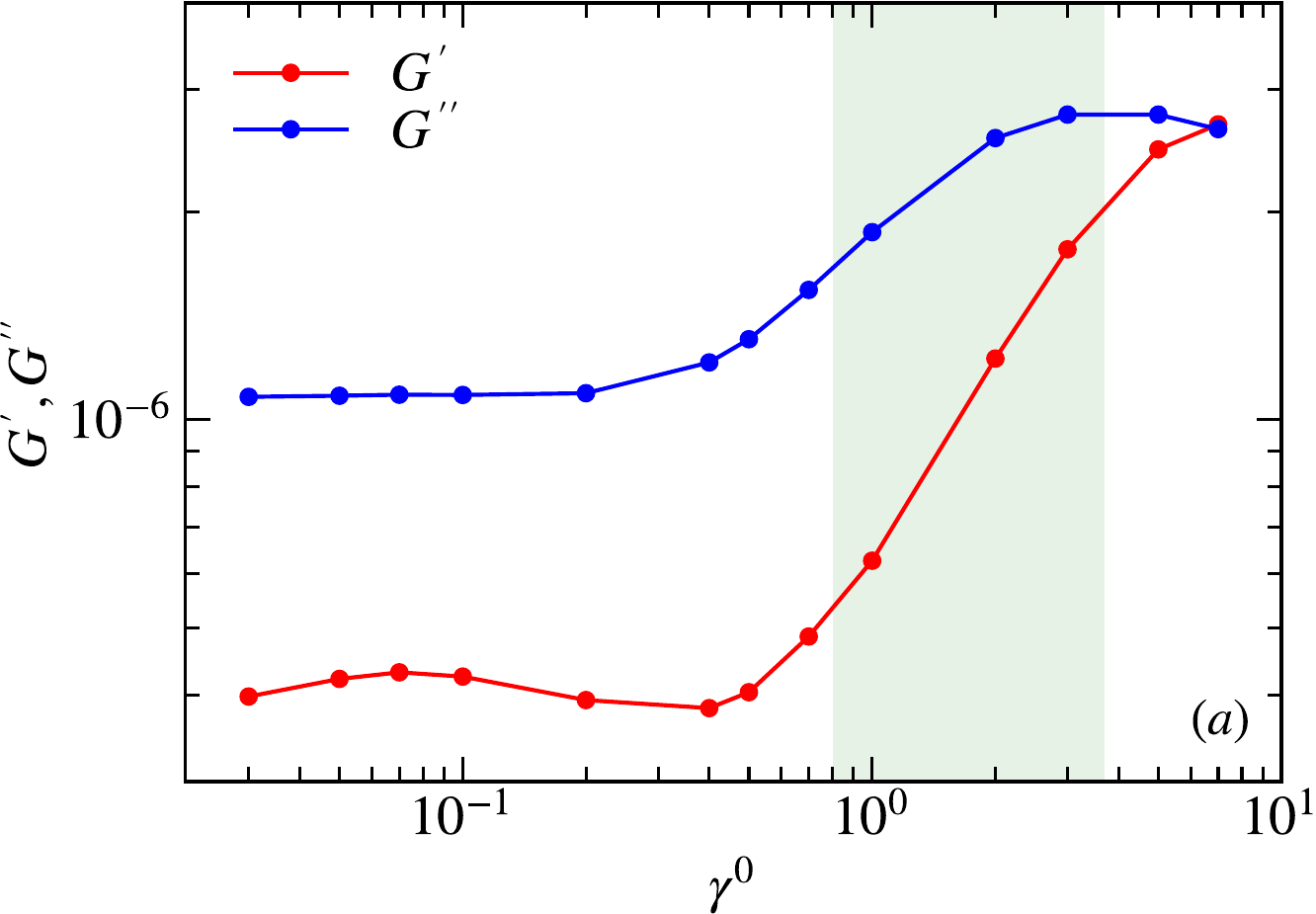}
  \includegraphics[width=\columnwidth]{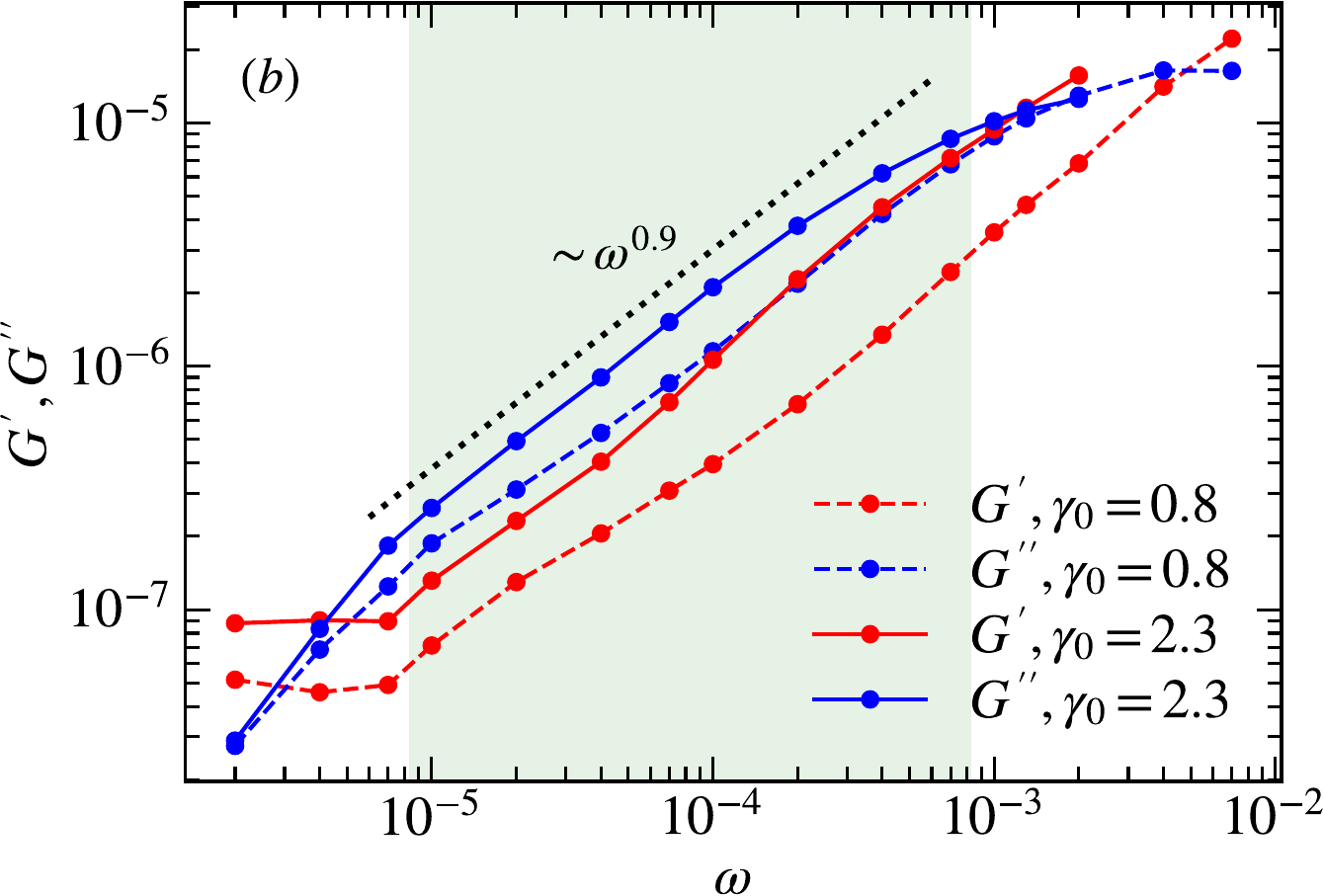}
  \caption{
(a) Influence of the squirmer active stress $\beta$ on the dynamic moduli of the polymer suspension, shown as an effective oscillatory-shear response at fixed frequency $\omega = 1.38\times10^{-4}\,\tau^{-1}$, corresponding to a swimming speed $U_{0}=3\times10^{-3}\,a\tau^{-1}$. Swimmers with $|\beta|\le 5$ generate  effective strain amplitudes $\gamma_{0}=0.8\text{--}3.7$ (green region). (b) Influence of the squirmer swimming speed $U_{0}$ on the dynamic moduli at fixed effective strain amplitudes $\gamma_{0}=0.8$ (neutral swimmer) and $\gamma_{0}=2.3$ (pusher/puller with $|\beta|=3$). The range $U_{0}=3\times10^{-4}\text{--}3\times10^{-2}\,a\tau^{-1}$ corresponds to effective oscillatory frequencies $\omega = 8.3\times10^{-6}\text{--}8.3\times10^{-4}\,\tau^{-1}$ (green region).}
  \label{osi}
\end{figure}

In Fig.~\ref{osi}(a), the influence of active stress on the viscoelastic response of a polymeric suspension containing $N_{p} = 50$ polymers of length $L_{p} = 240$ is presented. Squirmers swim at a fixed velocity $U_{0} = 0.003\, a\tau^{-1}$, corresponding to an effective oscillatory frequency $\omega = 1.38 \times 10^{-4}\,\tau^{-1}$. A broad linear viscoelastic regime is observed for $0.03 \leq \gamma_{0} \leq 0.5$. At larger strain amplitudes $\gamma_{0}=0.8 \text{--} 3.7$, corresponding to squirmers of $|\beta| \leq 5$, both $G'$ and $G''$ exhibit pronounced strain-thickening and cross near $\gamma_{0} \approx 7$. The system remains liquid-like, with large loss tangents $G''/G' \approx 1.4\text{--}3.2$, indicating that swimmer-induced shear is insufficient to generate a strong local elastic response; instead, viscous dissipation dominates across all active stresses. It is worth noting that the simulation height is chosen as $H = 38\,a$ in the oscillatory shear rheometry, but varying the box height while keeping the oscillation frequency fixed does not produce qualitative changes.

The dependence of dynamic moduli on swimming velocity for neutral swimmers and pushers/pullers with $|\beta| = 3$ is presented in Fig.~\ref{osi}(b). According to Eq.~\ref{strain_amp_s} and Fig.~S1, the corresponding strain amplitudes are $\gamma_{0} = 0.8$ and $\gamma_{0} = 2.3$, respectively. For $\gamma_{0} = 0.8$ (neutral swimmer), the moduli exhibit a power-law scaling $\sim \omega^{0.9}$ over $\omega = 7\times10^{-6} \text{--} 2\times10^{-3}\,\tau^{-1}$. The loss tangent $G''/G' \approx 1.9 \text{--} 3.1$ varies only weakly, indicating dominant viscous dissipation. A crossover near $\omega \approx 5 \times 10^{-3}\,\tau^{-1}$ signals the transition from the intermediate regime to the glassy regime, where polymers are largely immobilized and only local vibrational and rotational modes contribute. However, the physically relevant frequency window for swimmer-induced shear is $\omega = 8.3\times10^{-6}\text{--}8.3\times10^{-4}\,\tau^{-1}$, corresponding to $U_0 = 3\times10^{-4}\text{--}3\times10^{-2}\,a\tau^{-1}$, since very low or very high swimming velocities fall outside the reliable lattice Boltzmann resolution. Within this range, neutral-swimmer activity produces a strongly viscous response. For $\gamma_0 = 2.3$ ($|\beta| = 3$), $G''$ exhibits a similar power-law behavior in the intermediate window $\omega = 7\times10^{-6}\text{--}4\times10^{-4}\,\tau^{-1}$. In this regime, the loss tangent is $G''/G' \approx 1.4 \sim 2.2$, confirming viscous dominance. At $\omega = 1.1\times10^{-3}\,\tau^{-1}$, the two moduli become nearly identical, signaling the onset of the glassy transition. Similarly, both pullers and pushers induce a predominantly viscous polymeric response across the accessible frequency range.

\begin{figure}
  \centering
  \includegraphics[width=\columnwidth]{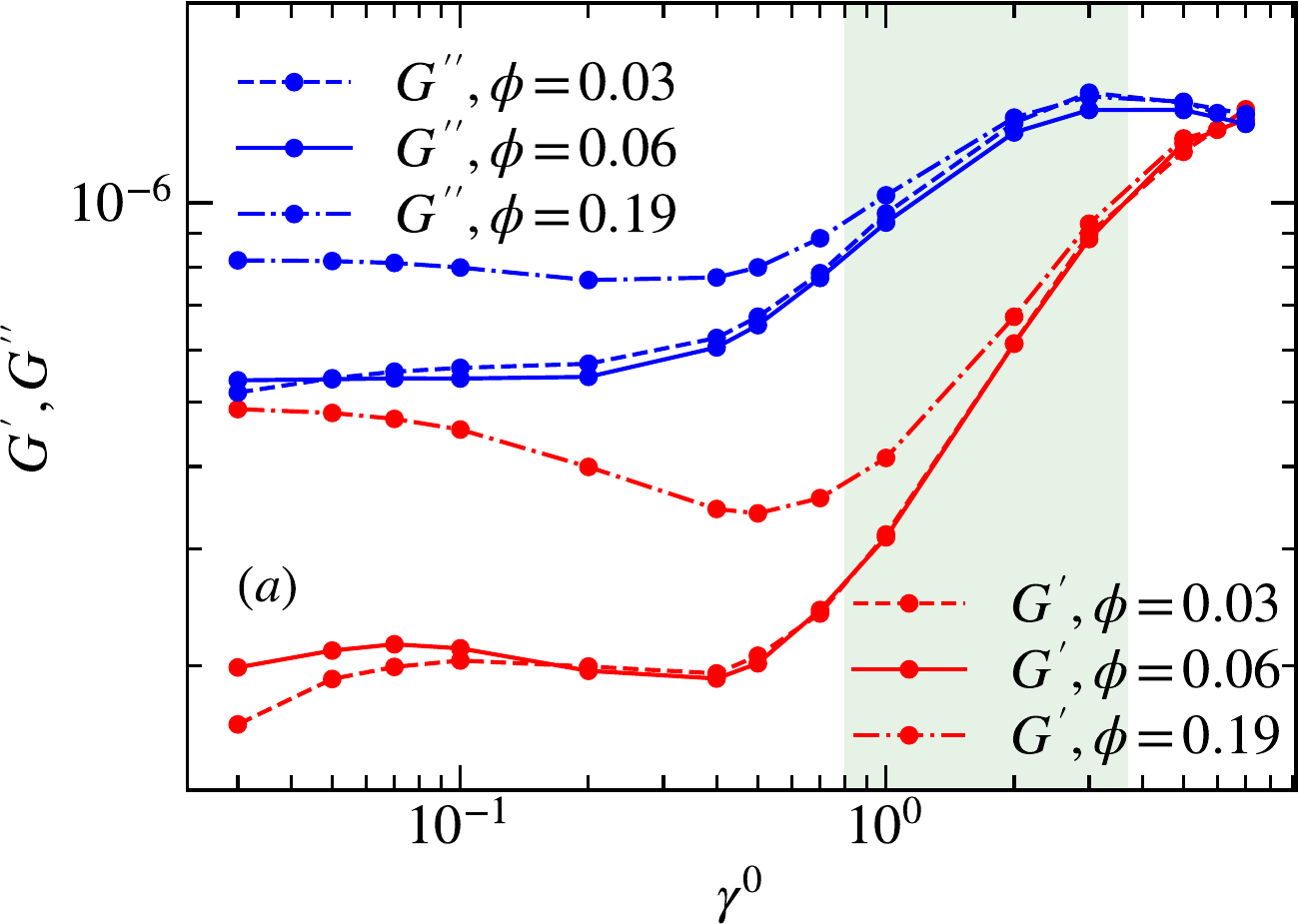}
  \includegraphics[width=\columnwidth]{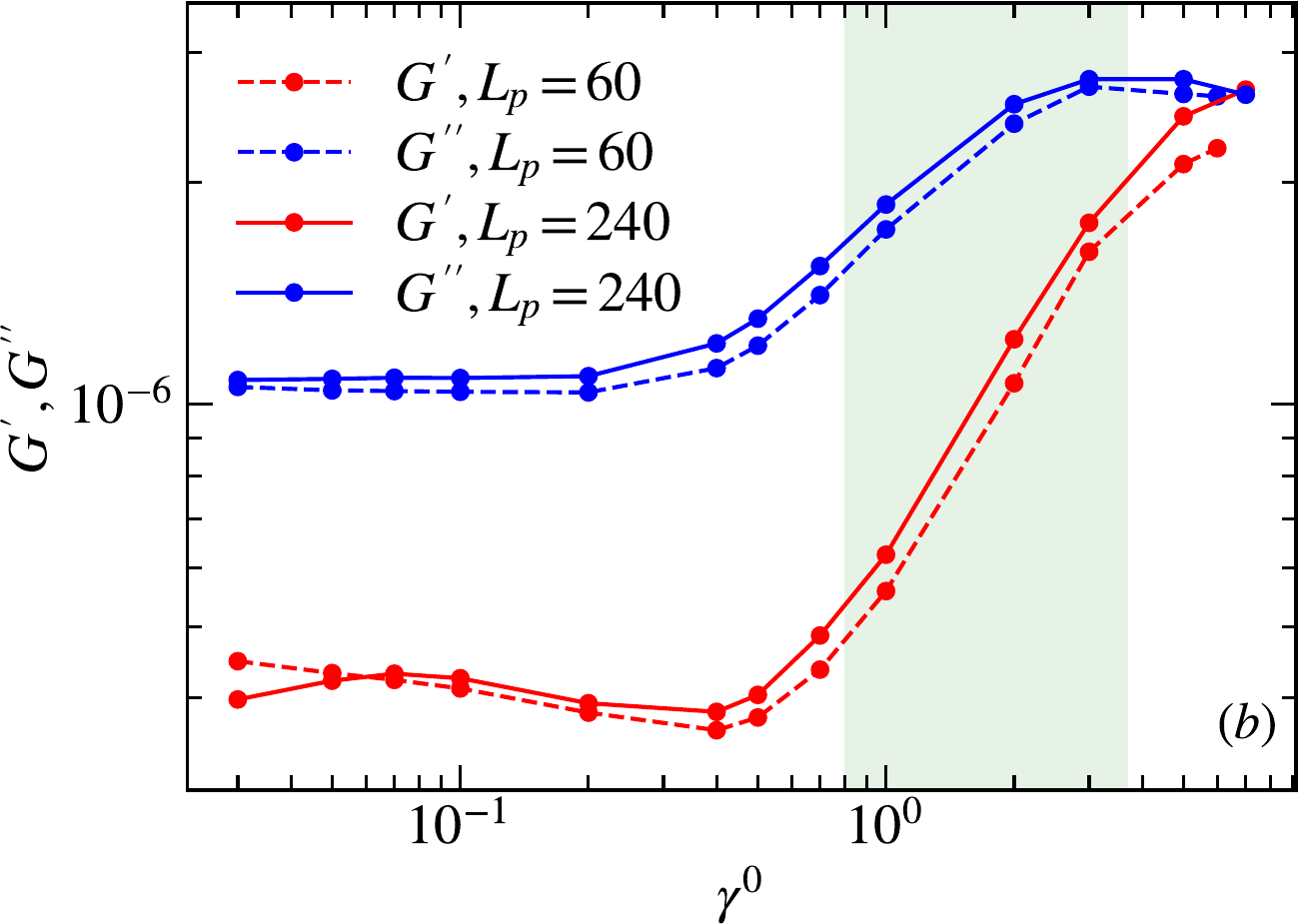}
    \caption{
    (a) Influence of polymer packing fraction $\phi$ on the dynamic moduli of polymer suspensions under shear induced by squirmer activity. The swimming velocity is fixed at $U_{0} = 0.003\,a\tau^{-1}$, corresponding to an effective oscillation frequency $\omega = 1.38 \times 10^{-4}\,\tau^{-1}$. For clarity of presentation, the dynamic moduli at $\phi = 0.06\,a^{-3}$ and $\phi = 0.19\,a^{-3}$ are rescaled by factors of $1/2$ and $1/6$, respectively. (b) Influence of polymer length $L_{p}$ on the dynamic moduli under the same conditions. Here, the polymer packing fraction $\phi= 0.06\,a^{-3}$ is fixed. Swimmers with $|\beta|\leq 5$ correspond to effective strain amplitude $\gamma_{0}=0.8 \text{--} 3.7$ (green region).}
  \label{osi_Np_Lp}
\end{figure}

The influence of the monomer packing concentration $\phi = L_{p}N_{p}/V$ on the dynamic moduli for squirmers with a fixed swimming speed $U_{0} = 0.003\,a\tau^{-1}$, corresponding to an effective oscillation frequency $\omega = 1.38 \times 10^{-4}\,\tau^{-1}$, is shown in Fig.~\ref{osi_Np_Lp}(a). For clarity of presentation, the dynamic moduli at $\phi = 0.06\,a^{-3}$ and $\phi = 0.19\,a^{-3}$ are rescaled by factors of $1/2$ and $1/6$, respectively. According to Eqs.~\ref{stress_xz} and \ref{moduli_def}, both dynamic moduli are expected to scale as $G', G'' \sim \phi^{2}$. However, since the FENE bond acts only between adjacent monomers and the repulsive potential is short-ranged, an asymptotic linear dependence of both moduli on $\phi$ is observed, while the overall shapes of the moduli remain largely preserved. This behavior is particularly evident for strain amplitudes $\gamma_{0} > 1$. For $\gamma_0 \le 1$, modest oscillations cannot generate substantial separations between polymers, and therefore the system at $\phi = 0.19\,a^{-3}$ exhibits enhanced rescaled storage and loss moduli, reflecting additional stress contributions from non-bonded monomer--monomer repulsions within densely compacted regions. For swimmers with $|\beta| \le 5$ ($\gamma_0 = 0.8\text{--}3.7$), the loss modulus consistently exceeds the storage modulus across all packing fractions, indicating a robustly viscous-dominated response to squirmer-induced shear.

The dependence of the dynamic moduli on polymer length is shown in Fig.~\ref{osi_Np_Lp}(b). Shortening the polymers leads to only a modest reduction in both moduli. Notably, $G'$ and $G''$ become separated near $\gamma_0 \approx 7$, indicating a delayed glass transition occurring at larger strain amplitudes. Across the relevant range $\gamma_0 = 0.8\text{--}3.7$, the loss tangent satisfies $G''/G' > 1$, confirming that the fluid response remains predominantly viscous.

To validate the analogy between swimmer-generated shear and that imposed in oscillatory shear rheometry, hydrodynamic simulations of a single squirmer embedded in a polymer solution were performed in our previous work under identical fluid conditions \cite{qi2025unravel}. The mapping between the representative squirmer parameters and the corresponding rheometric control variables is summarized in Table~\ref{tab_gamma_omega}. As shown in Fig.~\ref{osi}, all swimmers operate in the viscous-dominant regime, where elastic contributions are negligible. This behavior is consistent with our earlier results~\cite{qi2025unravel}, in which squirmers across a range of Reynolds numbers displayed up to three orders of magnitude enhancement in rotational diffusion. This pronounced effect arises from coupled mechanical and hydrodynamic interactions with surrounding polymers: the swimmer deforms nearby chains, producing restoring forces and torques through direct contacts and asymmetric local flows. In particular, polymers tend to wrap ahead of pushers and accumulate behind pullers. Here, no apparent elastic effect was observed. These interactions intensify with polymer concentration, yielding maximal rotational diffusion at $\phi = 0.19 \, a^{-3}$ for pullers~\cite{qi2025unravel}. As shown in Fig.~\ref{osi_Np_Lp}(a), viscous dissipation dominates across all concentrations, in agreement with the hydrodynamic simulations. Likewise, polymer length has only a minor influence, as reflected in Fig.~\ref{osi_Np_Lp}(b), where the dynamic moduli for different chain lengths remain closely clustered, again consistent with Ref.~\cite{qi2025unravel}. We note, however, that modifying polymer properties---such as bond extensibility or bending rigidity---may shift the system into a regime in which elastic effects dominate under squirmer-induced shear.

\setlength{\tabcolsep}{0.17em}
\begin{table}[t]
    \caption{Mapping from the squirmer to oscillatory shear rheometry. The squirmer radius is $R = 3a$, and the characteristic length scale is $\xi = 38a$.}
    \begin{tabular}{ c c | c c| c c c }
        ${\rm U_0} (\times 10^{-3})$ & $\beta$ & $\omega (\times 10^{-4})$ & $\gamma_0$ & $G' (\times 10^{-6})$ & $G'' (\times 10^{-6})$ & $G''/G'$ \\ \hline
         3.3 &  -5 &  1.38 &  3.7 & 2.0 & 2.7 & 1.35 \\
         3.3 &  -2 &  1.38 &  1.6 & 1.6 & 2.6 & 1.63 \\
         3.3 &  0  &  1.38 &  0.8 & 0.5 & 1.6 & 3.20 \\
         3.3 &  2  &  1.38 &  1.6 & 1.6 & 2.6 & 1.63 \\
         3.3 &  5  &  1.38 &  3.7 & 2.0 & 2.7 & 1.35 
    \end{tabular}
    \label{tab_gamma_omega}
\end{table}

To summarize, we propose a strategy that maps the classical oscillatory shear rheometry framework onto the effective shear generated by microswimmers, thereby enabling quantitative evaluation of the viscoelastic response of fluids directly elicited by swimmer activity. The mapping relies on identifying a characteristic length scale $\xi$ at which the swimmer’s flow effectively vanishes. In the rheometric analogy, this distance acts as a static imaginary wall in the vertical direction, while laterally it serves as a quarter-wavelength of the oscillation. The key step is establishing a correspondence between the mean shear rate induced by swimming and that generated by an oscillating plate. Once this mapping is achieved, the swimming speed and active stress can be translated into an equivalent oscillatory frequency and strain amplitude, enabling standard oscillatory shear rheometry to probe swimmer-induced viscoelasticity. This generic protocol thus provides a pathway to quantify activity--induced rheology and can be naturally extended to active microrheology, where an analogous mapping can be constructed between the mean shear generated by driven probe particles and that induced by microswimmers.

\section{Acknowledgements}

The authors thank Dr.~Marco De Corato for valuable suggestions. This work was supported by the Swiss National Science Foundation through the program \textit{Computational Modeling at CECAM: Challenges in the Foundations and Modeling of Systems Far from Equilibrium} (200021\_175719), and by the National Natural Science Foundation of China (Nos.~12304257 and~12574237). Computing resources were provided by the Piz~Daint supercomputer at the Swiss National Supercomputing Centre (CSCS).

\section{Data availability}
The data that support the findings of this article are not publicly available. The data are available from the authors upon reasonable request.


\bibliographystyle{apsrev4-2}
\bibliography{squ_poly_osi}

\section{End Matter}

\subsection{Characteristic length of a squirmer}
According to Eq.~\ref{flow_field_squ_full}, The squared flow field is  
\begin{align}
    v^2(r, \theta) &= \frac{B_1^2 R^6}{3 r^6} \cos^2 \theta + \frac{B_1^2 R^6}{9 r^6} + \frac{9 B_2^2 R^4}{4 r^4} \cos^4 \theta \notag \\
    & - \frac{3B_2^2 R^4}{2 r^4} \cos^2 \theta + \frac{B_2^2 R^4}{4 r^4} - 2 B_1 B_2 \frac{R^5}{r^5} \cos^3 \theta \notag \\
    & + \frac{2}{3} B_1 B_2 \frac{R^5}{r^5} \cos \theta. \label{v_sqr}
\end{align}
The mean flow field at a distance $r$ from the center of mass of a squirmer is thus obtained by integrating Eq.~ \ref{v_sqr} over the polar angle $\theta$ and the azimuthal angle $\phi$ as
\begin{equation}
    \langle v^2(r) \rangle = \frac{1}{4\pi} \int v^2 \sin \theta d\theta d\phi = \frac{2 B_1^2 R^6}{9 r^6} + \frac{B_2^2 R^4}{5r^4}. \label{v_sqr_avg} 
\end{equation}
The flow field decay can thus be estimated as 
\begin{equation}
    \lambda=\sqrt{\frac{\langle v^2(r) \rangle}{\langle v^2(R) \rangle}}=\left(\frac{2 R^6}{9 r^6} + \frac{\beta^2 R^4}{5 r^4}\right)^{1/2} \left(\frac{2}{9} + \frac{\beta^2}{5}\right)^{-1/2}, \label{decay strength}
\end{equation}
which decays to approximately $0.5\%$ at a separation $\xi = 38\,a$ for most squirmers with $|\beta| \le 5$. According to Eq.~\ref{v_sqr_avg}, the mean flow velocity at this distance is $v_b = 2.8 \times 10^{-5}\, a/\tau$ for a representative squirmer with active stress $\beta = 2$ and swimming speed $U_0 = 0.003\, a/\tau$. For an inertialess monomer in our system, the renormalized hydrodynamic radius is $a_R = 0.24\,a$, which yields a translational diffusion coefficient $D_t = k_B T/6\pi\eta a_R = 4.5 \times 10^{-7}\, a^2/\tau$ (see SM for parameters). Thus, the P\'eclet number for a free monomer suspended in the fluid is $\mathrm{Pe}_m = 2a_R v_b/D_t = 30$. However, since monomers are constrained by strong bond potentials within the polymer chain, such a modest P\'eclet number cannot induce appreciable flow-driven motion. Therefore, we choose $\xi = 38\,a$ as the characteristic length scale of the squirmer.

\subsection{Derivation for the shear rate due to swimming}

The shear rate induced by squirmer activity in three dimensions can be estimated as 
\begin{align}
    \notag \langle \dot{\gamma}^2_s \rangle &=\langle 2 \mathbf{D}:\mathbf{D} \rangle \\
    &= \langle 2D_{rr}^2 + 2D_{\theta\theta}^2 + 2D_{\phi\phi}^2 + 4D_{r\theta}^2 + 4D_{r\phi}^2 + 4D_{\theta\phi}^2 \rangle, \label{mean_shear_rate_s_deri}
\end{align}
where $\mathbf{D}$ denotes the strain-rate tensor with components defined by $D_{ij} = \frac{1}{2}\left( (\nabla\mathbf{v})_{ij} + (\nabla\mathbf{v})_{ji} \right)$ \cite{ladisa2005circumferential}. Here, $\mathbf{v}$ is the flow field generated by squirming, described in Eq.~\ref{flow_field_squ_full}. The tensor components in spherical coordinates are obtained from the general expression for the strain-rate tensor in an orthogonal curvilinear coordinate system, with scale factors $h_{r} = 1$, $h_{\theta} = r$, and $h_{\phi} = r \sin\theta$. Specifically, the components are 
\begin{align}
    D_{rr} &= \frac{\partial v_r}{\partial r} \\
    D_{\theta\theta} &= \frac{1}{r} \frac{\partial v_\theta}{\partial \theta} + \frac{v_r}{r} \\
    D_{\phi\phi} &= \frac{1}{r \sin\theta} \frac{\partial v_\phi}{\partial \phi} + \frac{v_r}{r} + \frac{v_\theta \cot\theta}{r}\\
    D_{r\theta} &= D_{\theta r} = \frac{1}{2} \left( \frac{\partial v_\theta}{\partial r} - \frac{v_\theta}{r} + \frac{1}{r} \frac{\partial v_r}{\partial \theta} \right) \\
    D_{r\phi} &= D_{\phi r} = \frac{1}{2} \left( \frac{1}{r \sin\theta} \frac{\partial v_r}{\partial \phi} + \frac{\partial v_\phi}{\partial r} - \frac{v_\phi}{r} \right) \\
    D_{\theta\phi} &= D_{\phi\theta} = \frac{1}{2} \left( \frac{1}{r} \frac{\partial v_\phi}{\partial \theta} - \frac{v_\phi \cot\theta}{r} + \frac{1}{r \sin\theta} \frac{\partial v_\theta}{\partial \phi} \right). \label{D_comp}
\end{align}
The velocity components are explicitly written as
\begin{align}
    v_r &=  \frac{2}{3} B_1 \frac{R^3}{r^3} \cos \theta - \frac{B_2 R^2}{2r^2} (3 \cos^2 \theta - 1)\\
    v_\theta &= \frac{B_1 R^3}{3r^3} \sin \theta \\
    v_\phi &= 0. \label{v_comp}
\end{align}
Correspondingly, the velocity gradients are calculated as
\begin{align}
    \frac{\partial v_r}{\partial r} &=  -\frac{2B_1 R^3}{r^4} \cos \theta + \frac{B_2 R^2}{r^3} (3 \cos^2 \theta - 1)\\
    \frac{\partial v_r}{\partial \theta} &= \frac{3B_2 R^2}{r^2}\cos \theta \sin \theta - \frac{2B_1 R^3}{3r^3} \sin \theta \\
    \frac{\partial v_\theta}{\partial r} &= -\frac{B_1 R^3}{r^4} \sin \theta\\
    \frac{\partial v_\theta}{\partial \theta} &= \frac{B_1 R^3}{3r^3} \cos \theta \\
    \frac{\partial v_r}{\partial \phi} &= \frac{\partial v_\theta}{\partial \phi} = \frac{\partial v_\phi}{\partial r} = \frac{\partial v_\phi}{\partial \theta} = \frac{\partial v_\phi}{\partial \phi}=0. \label{v_grad}
\end{align}
Therefore, the components of the strain-rate tensor in spherical coordinates can be explicitly expressed as
\begin{align}
    D_{rr} &=-\frac{2B_1 R^3}{r^4} \cos \theta + \frac{B_2 R^2}{r^3} (3 \cos^2 \theta - 1) \label{Drr} \\
    D_{\theta\theta} &= \frac{B_1 R^3}{r^4} \cos \theta - \frac{B_2 R^2}{2r^3} (3 \cos^2 \theta - 1) \\
    D_{\phi\phi} &= \frac{B_1 R^3}{r^4} \cos \theta - \frac{3B_2 R^2}{2r^3} \cos^2 \theta + \frac{B_2 R^2}{2r^3} \\
    D_{r\theta} &= D_{\theta r} = \frac{1}{2} \left( \frac{3B_2 R^2}{r^3} \cos \theta \sin \theta - \frac{2B_1 R^3}{r^4} \sin \theta \right) \\
    D_{r\phi} &= D_{\phi r} = D_{\theta\phi} = D_{\phi\theta} = 0. \label{Drphi} 
\end{align}

By inserting Eqs.~\ref{Drr}--\ref{Drphi} into Eq.~\ref{mean_shear_rate_s_deri}, the mean shear rate due to swimming is obtained as 
\begin{widetext}
\begin{align}
    \langle \dot{\gamma}^2_s \rangle &= \left\langle \frac{27B_2^2 R^4}{r^6} \cos^4 \theta - \frac{36B_1 B_2 R^5}{r^7} \cos^3 \theta + \left( \frac{12B^2_1 R^6}{r^8} - \frac{18B^2_2R^4}{r^6} \right) \cos^2 \theta +\frac{3B_2^2 R^4}{r^6} \right. \notag \\
    &\quad \left. +\frac{12B_1 B_2 R^5}{r^7} \cos \theta + \frac{4B_1^2 R^6}{r^8} \sin^2 \theta + \frac{9B_2^2R^4}{r^6} \cos^2 \theta \sin^2 \theta - \frac{12B_1 B_2 R^5}{r^7} \cos \theta \sin^2 \theta \right\rangle \notag \\
    &= \frac{3}{4\pi (\xi^3 - R^3)} \left[ \left( \frac{1}{R^3} - \frac{1}{\xi^3} \right) \frac{24\pi}{5} B_2^2 R^4 +  \left( \frac{1}{R^5} - \frac{1}{\xi^5} \right)\frac{80}{15} \pi B_1^2 R^6 \right] \notag \\
    &= \frac{R B_1^2 }{\xi^3 - R^3} \left[ \left( 1 - \left( \frac{R}{\xi} \right)^3 \right)\frac{18}{5} \beta^2 + 4\left( 1 - \left( \frac{R}{\xi} \right)^5 \right) \right] \notag \\
    & \approx \frac{R B_1^2 }{\xi^3} \left[ \frac{18}{5}\beta^2 + 4 \right] = \frac{9R U_0^2 }{4\xi^3} \left[ \frac{18}{5}\beta^2 + 4 \right], \label{gamma_s_derive}
\end{align}
\end{widetext}
where $B_1=3/2U_0$ is utilized and the averaging is carried out over the entire spherical shell between $R$ and $\xi$, under the assumption $\xi \gg R$.

\end{document}